# Application of the analytical methods REM/EDX, AES and SNMS to a chlorine induced aluminium corrosion

Uwe Scheithauer

*ZPL 1 TW 45, Otto-Hahn-Ring 6, 8000 München 83, Fed. Rep. of Germany*[1]

## Keywords

## Abstract

Scanning electron microscopy (SEM) with energy dispersive X-ray detection (EDX), Auger electron spectroscopy (AES) and sputtered neutral mass spectrometry (SNMS) have been used to characterize a chlorine induced corrosion of an aluminium metallisation. SEM/EDX detects the characteristic X-rays that are emitted from the first few micrometers beneath the specimen's surface after inner shell ionisation by the primary electrons. AES detects the alternatively ejected Auger electrons that are generated within the topmost atomic layers of the sample. To obtain elemental concentration depth profiles, the surface layers are removed by ion sputtering. Whereas AES detects the composition of the remaining surface, SNMS measures sputtered fluxes and does not suffer from preferential sputtering. As demonstrated by the example of a chlorine induced aluminium corrosion, these analytical methods are complementary with respect to quantification, chemical information and information depth. Only by simultaneous use measuring artefacts are detectable and able to be excluded from interpretation.

## 1. Introduction

In high power devices thin aluminium layers are used as large scale contact metallisation. Currents with densities of approximately 100 A/cm$^2$ flow perpendicular across these layers. The mechanical deformability of aluminium provides for a good electrical contact to a nickel electrode of some centimetres diameter that is pressed onto the metallisation. After testing these devices in industrial atmosphere some long time failures of the aluminium layers due to high contact resistances have been observed. To analyse this problem, the surface of the aluminium metallisation was characterized by SEM/EDX [1], AES [2, 3] and SNMS [4, 5, 6].

## 2. Experimental Results

### SEM/EDX

Energy dispersive-ray spectra have been taken in a scanning electron microscope (SEM) Cambridge S 200 equipped with a Si(Li) detector. Due to the Be entrance window and its absorption of X-rays at the low energy side, only K lines of elements to a minimum of Z=11 (Na) can be analysed. The detector crystal is mounted rectangularly with respect to the primary beam. The sample to be analysed must be tilted to perform a measurement.

Fig. 1 shows two examples of EDX spectra. Silicon, sulphur, chlorine and nickel (abrasion of the contact electrode) are detected in addition to aluminium. To obtain a higher surface sensitivity, the sample is tilted towards a shallow angle of incidence and the primary beam energy $E_0$ is changed for the recording of the spectrum shown at the bottom of fig. 1. Lowering $E_0$ from 20 kV to 8 kV reduces the penetration depth of the primary electrons. A calculation, performed with the ionization energy $E_c(Cl)$ of the chlorine K shell and a specimen density of 2.7 g/cm$^3$, results in a reduction of the typical depth parameter of x-ray generation by a factor of approximately 5 [1]. Even in this case of low energy ($E_0$ = 8 kV) the parameter is in the order of 0.5 μm. Due to a larger tilt angle of the sample surface with respect to the primary beam an increase in interaction of the electrons with atoms takes place just beneath the surface. This also creates a greater fraction of the X-ray spectrum to monitor the composition of the uppermost layers. The surface roughness restricts the tilt angle to a maximum value of 75°. Both changes of the experimental parameters, the increasing tilt angle and the lower voltage, result in a higher sulphur and chlorine count rate relative to the aluminium signal.

To demonstrate this in more detail, fig. 2 shows the dependence of the normalized chlorine X-ray count rate as a function of primary energy and tilt angle. At approximately 7 kV primary energy, at an overvoltage ratio $E_0/E_c(Cl) = 2.5$, a maximum value of the chlorine count rate is reached. At the lower energy side the signal decreases due to the strongly reduced ionisation cross section at low overvoltage ratios [2, 7]. At the high energy side the signal also decreases. The ionisation probability is a slowly varying function at high overvoltage ratios between 2.5 and 8 [2, 7]. Because of this, one would expect no significant energy and tilt angle dependence for overvoltage ratios greater than 2.5 for a depth independent chlorine concentration. Therefore the measurements of energy dispersive x-ray spectra show that chlorine is enriched at the sample surface. Analogous measurements of the energy and tilt angle dependence of the normalized sulphur x-ray count rate have similar results.

To estimate the exact depth profile distribution of these contaminations, analytical methods with higher depth resolution must be used.

## AES

For the AES analysis a scanning Auger microprobe type PHI 660 has been used. The system is equipped with a cylindrical-mirror electron energy analyser and an integral, co-axially mounted $LaB_6$ electron gun. For sputter-etching of the surface and depth profiling a differential pumped argon ion gun is installed. If measurements are performed under normal working conditions the sample is inclined by 30° with respect to the axis of the cylindrical-mirror analyser. In this case the argon ions hit the surface at an angle of 55° relative to the surface normal.

Fig. 3 shows AES spectra that have been recorded in different sputter depths. The sample was analysed within an area of approximately $10^3$ $\mu m^2$. Aluminium, silicon, nickel, oxygen, carbon and chlorine are detected. The details of the aluminium KLL-Auger line shape should be emphasized. At the surface down to approximately 100...150 nm sputter depth the line shape has non-metallic features [8]. In a sputter depth of 170 nm the plasmon excitations indicate metallic aluminium. This progress corresponds to the development of the oxygen concentration decreasing with depth. Within the first 100...150 nm the aluminium is oxidized.

From the measured peak to peak signals of the differentiated AES spectra atomic concentration profiles are derivated [9] by means of pure element standard sensitivity factors [8]. Fig. 4 shows the estimated atomic concentration profiles. Due to vast differences of sensitivity factors the detection limits of the elements differ.

To calibrate the depth scale of the sputter profile, the sputter yield of the sample was related to that of a reference material, a $SiO_2$ layer of known thickness, used to determine the sputter erosion of the ion gun. Here for the sample with a composition similar to $Al_2O_3$ a sputter yield of 1/5 of the $SiO_2$ sputter yield was selected from the different values reported in literature for $Al_2O_3$ [10]. Using these parameters, the thickness of the insulating surface layer is in the order of that estimated by measuring the electrical break-down voltage of this layer.

Near the surface silicon and nickel are detected, both signals decreasing with depth. A carbon contamination is present within the whole $Al_2O_3$ layer. Possible contribution to this signal may be preferential sputtering in a sense of low carbon sputter rates or recontamination from vacuum. Chlorine, sulphur and in some sputter depths fluorine are present just above the detection limit.

Since Auger electron spectroscopy analyses the composition of the surface after ion etching, preferential sputtering effects, which may cause deteriorations in the composition, have to be taken into account. Also modifications due to the primary electron beam may have some influence on the surface composition. All this is hardly to be calculated for a sample of unknown composition.

## SNMS

Whereas AES detects the composition of the surface remaining after sputtering, SNMS measures the sputtered fluxes and thereby does not suffer from preferential sputtering if steady state is reached. During a SNMS analysis, a low pressure ($10^{-4}$ mbar range) rf-coupled argon plasma is established in front of the sample. The sample is biased negatively to the plasma. Argon ions are extracted and erode the surface within an area of a minimal diameter of 2 mm. The sputtered neutrals are emitted into the plasma and are post-ionized by electron impact. These ions are detected by a quadrupole mass spectrometer. Since erosion and postionization are decoupled, the ionization process is not influenced by matrix effects.

Standardless quantitative depth profile analysis of samples of unknown composition is well established.

Fig. 5 shows the results of a SNMS depth profile analysis. To estimate the thickness of the insulating layer, the decrease of the oxygen signal can be measured. It drops from 50 at% just beneath the surface to half of the value in a depth of approximately 150 nm. The high sulphur and chlorine signals are one order of magnitude above the concentrations estimated by AES depth profile measurement. Because of lower detection limits of SNMS nickel is found up to greater depth inside the sample.

## 3 Conclusions

SNMS and AES analyse the composition of the surface within the first few atomic layers. Elemental concentration depth profiles were obtained by sputter erosion of the surface. To the characteristic X-rays measured with SEM/EDX atoms of the first few micrometres contribute to the signal. By variation of the primary energy and the sample tilt angle the surface sensitivity was enhanced within certain limitations. SEM/EDX gives a quick first glance to the problem to be analysed.

SNMS detects a chlorine and sulphur concentration inside the insulating layer approximately a factor of 10 above that measured with AES sputter depth profiling. The SNMS results reproduce the real situation due to the lack of preferential sputtering and/or modifications of the surface composition by a primary electron beam. The depth resolution of the SNMS depth profile is poorer than that of AES. Considering the larger analysis area of the SNMS instrument this gives a hint to a rough interface between the 150 nm thick oxide surface layer and aluminium.

AES gives some additional chemical information about the insulating layer. In a depth of approximately 150 nm the features of the aluminium KLL Auger signal point out a change from non-metallic to metallic aluminium. This progress and the oxygen concentration decreasing at this depth indicate that aluminium is oxidized within the first 150 nm of the sample. Considering the high chlorine concentration of the insulating surface layer, a reaction mechanism is probable, where aluminium is catalytically converted into its oxide if moisture is present [11]. The main steps of this reaction may be expressed as follows:

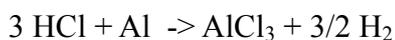
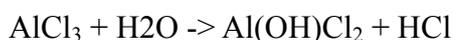

$$3\ HCl + Al \rightarrow AlCl_3 + 3/2\ H_2$$
$$AlCl_3 + H_2O \rightarrow Al(OH)Cl_2 + HCl$$

If an insulating surface layer has been built up, the electrical resistance and thereby the thermal power dissipation of the aluminium metallisation increases. The device can be destroyed by thermal stress or melting.

The analytical capabilities of SEM/EDX, AES and SMNS have been demonstrated by the example of a corroded aluminium metallisation. Only by simultaneous use of the different methods measurement artefacts can be detected and are able to be excluded from interpretation.

## Acknowledgement

The SNMS measurements have been performed at Leybold AG, Cologne, FRG.

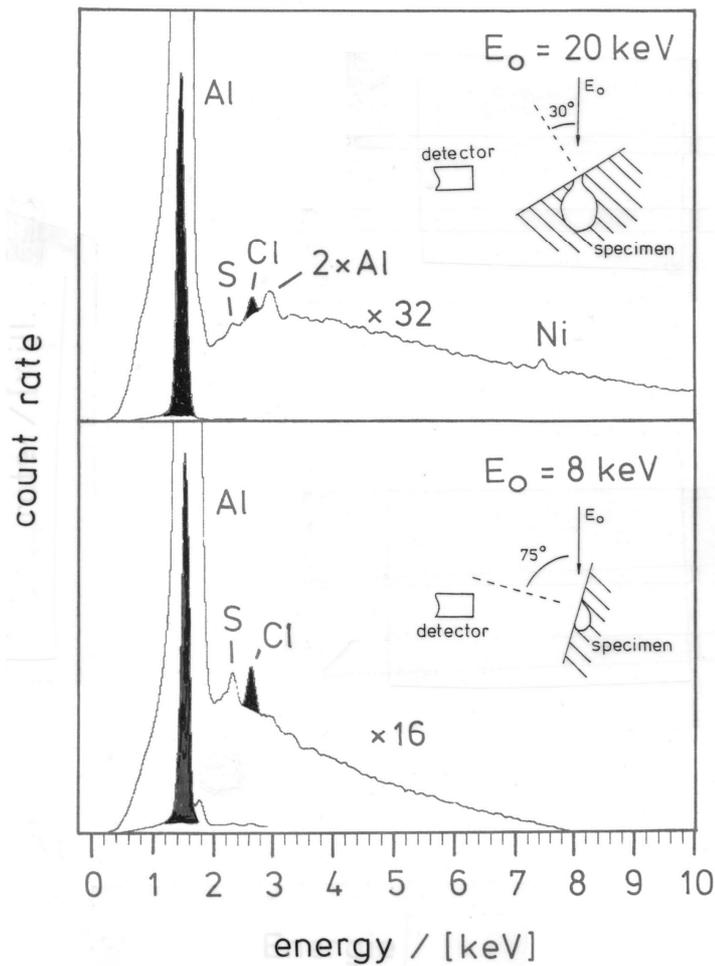

Fig. 1: EDX spectra. By lowering the primary energy $E_0$ and increasing the tilt angle of the specimen a higher surface sensitivity was obtained.

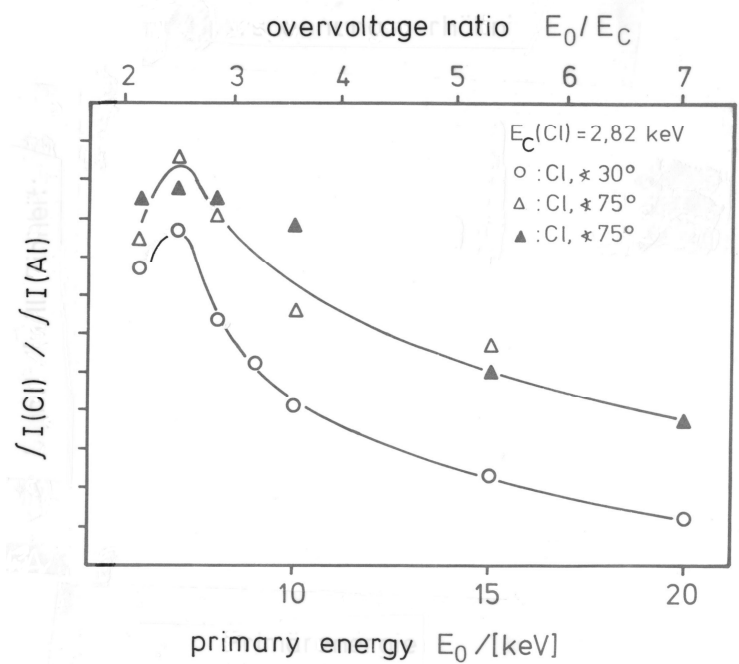

Fig. 2: normalized chlorine count rate as function of primary energy $E_0$ and tilt angle
○: Cl 30° tilt, ▲: Cl 70° tilt, Δ: Cl 70° tilt

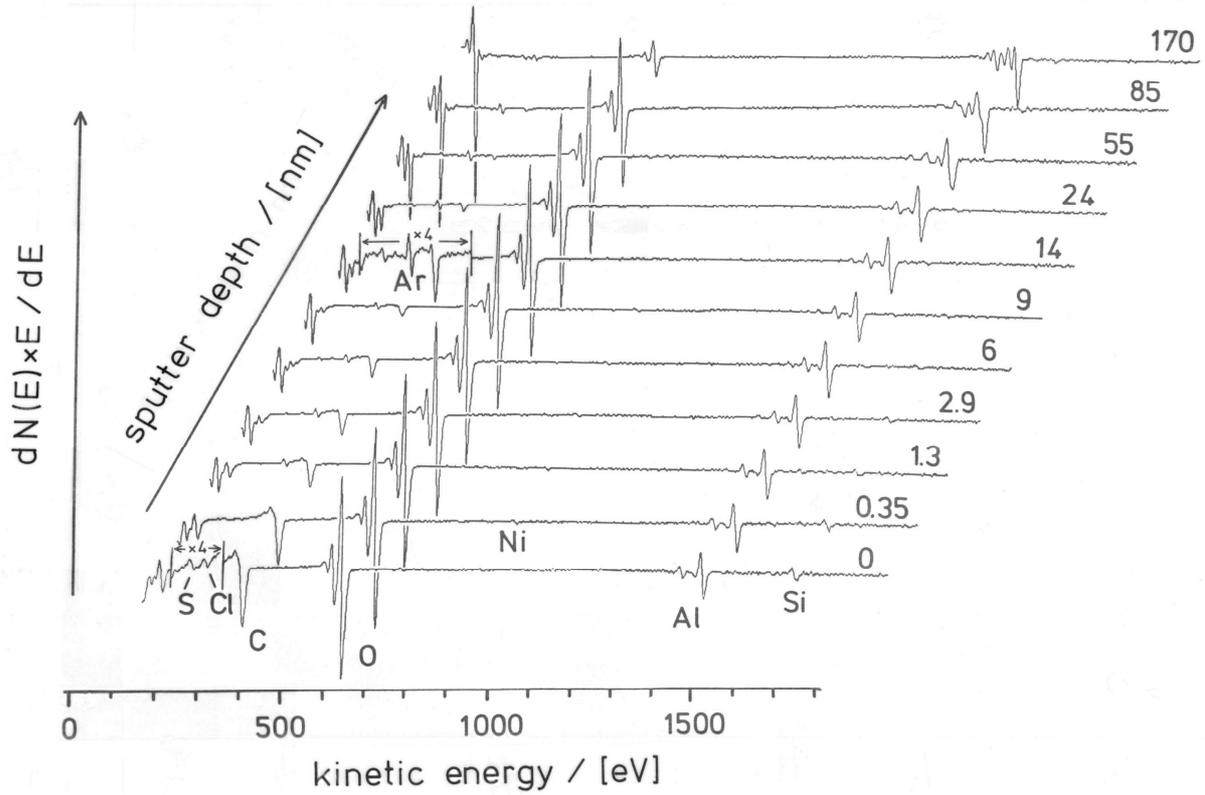

Fig. 3: AES spectra, recorded at different sputter depths

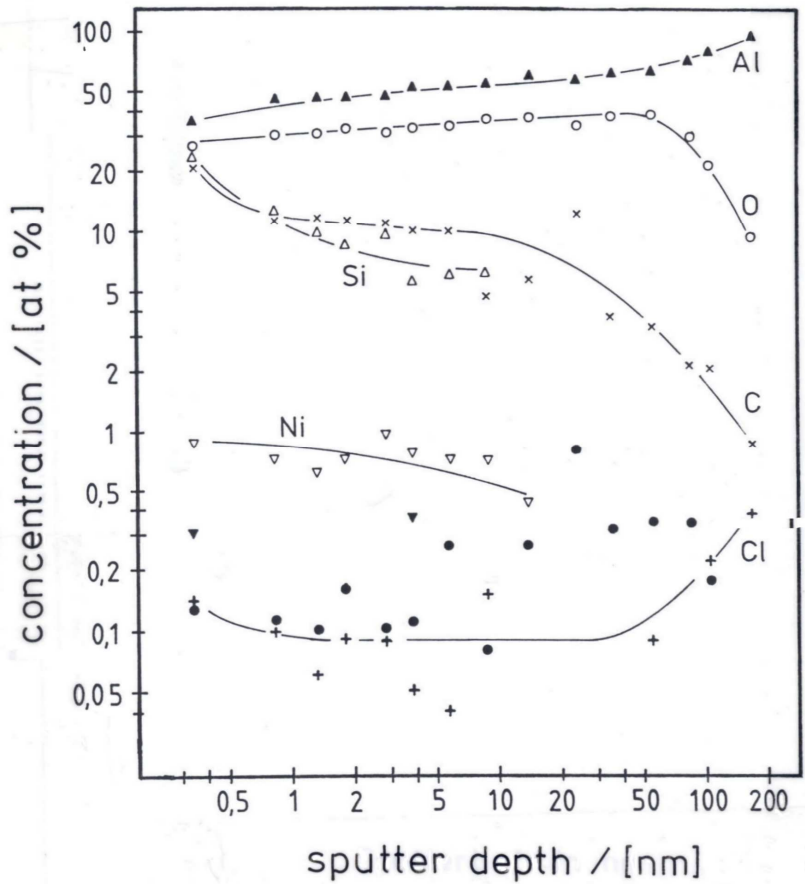

Fig. 4: AES sputter depth profile, Cl: + , S: ● , F: ▼

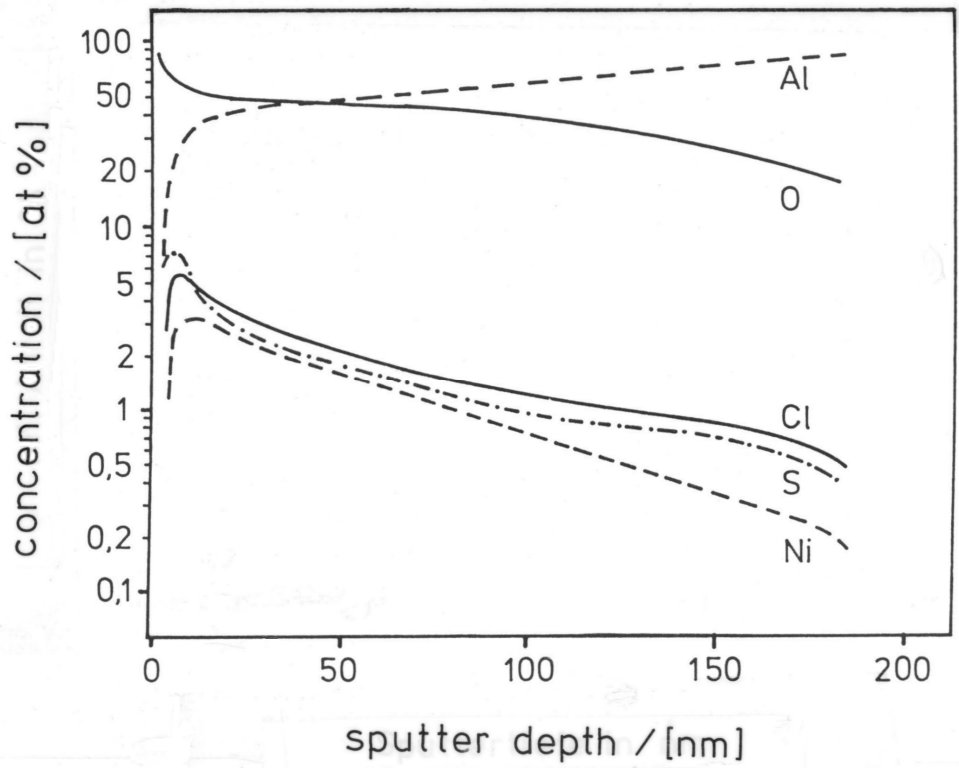

Fig. 5:   SNMS depth profile